# INTEGRATION OF THE VIMOS CONTROL SYSTEM

D. Mancini, P. Schipani, O. Caputi, G. Mancini, M. Brescia
Osservatorio Astronomico di Capodimonte, Napoli, Italy

Abstract

The VIRMOS consortium of French and Italian Institutes (PI: O. Le Fèvre, co-PI: G. Vettolani) is manufacturing two wide field imaging multi-object spectrographs for the European Southern Observatory Very Large Telescope (VLT), with emphasis on the ability to carry over spectroscopic surveys of large numbers of sources: the VIsible Multi-Object Spectrograph, VIMOS, and the Near InfraRed Multi-Object Spectrograph, NIRMOS.

VIMOS is being installed at the Nasmyth focus of the third Unit Telescope of the VLT at Mount Paranal in Chile, after a period of pre-integration in Europe at the Observatoire de Haute Provence. There are 52 motors to be controlled in parallel in the spectrograph, making VIMOS a complex machine to be handled. This paper will focus on the description of the control system, designed in the ESO VLT standard control concepts, and on some integration issues and problem solving strategies.

## 1  VIRMOS PROJECT

The need to collect large samples of objects has become a strong driver in the development of astronomical instrumentation. This is particularly evident for the observation of large samples of galaxies out to very large distances for which tens of thousands of objects are necessary to reach the accuracy needed in the measurement of fundamental astrophysical parameters. Multi-slit spectrographs allow to extract the source signal at much fainter levels than in multi-fiber spectrographs and are therefore now routinely used to reach the faintest magnitudes on several major telescopes.

In 1995, ESO launched a competitive call for proposal to study the feasibility of a Visible and/or Infrared Multi-Object Spectrograph, (VIRMOS), for the VLT. The VIRMOS consortium was awarded a one year contract in parallel with the Australis consortium of Australian institutes.

In October 1996, the VIRMOS proposal was selected by ESO. The project is now proceeding along a fast-track approach, with first light in early 2002 for VIMOS, the first to become operational. VIMOS and NIRMOS instruments allow for a large multiplex gain over a wide field, over the 0.37 to 1.8 μm domain. The unique multiplex gain allows to obtain spectra of up to 840 objects simultaneously with VIMOS, and up to 170 with NIRMOS (10 arcsec slits). An integral field spectroscopy mode with more than 6400 fibers coupled to micro-lenses will be available for VIMOS, covering a 1x1 arcmin² field. VIMOS and NIRMOS will cover up to 4x 8x8 arcmin^2. The instrument is divided into four channels. Each channel is in practice an imaging spectrograph provided with a field lens, a collimator, grisms or filter and a camera, coupled to a 2048x4096 pixel CCD for VIMOS, and a 2048x2048 HgCdTe Rockwell array for NIRMOS.

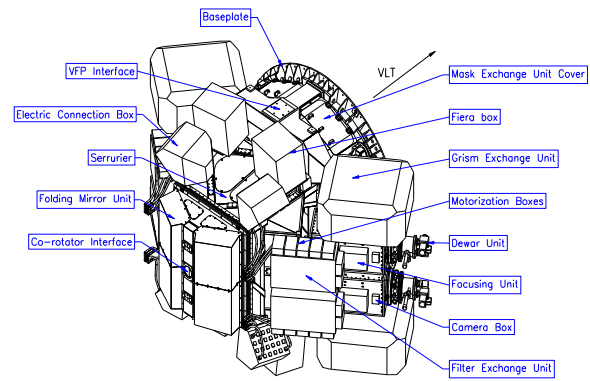

Figure 1: VIMOS spectrograph, CAD view

## 2  INSTRUMENT CONTROL

### 2.1  Subsystems

The VIMOS instrument can operate in three different modes:

- Multi-Object Spectroscopy mode
- Imaging mode
- Integral Field Unit (IFU) mode

This versatility implies a quite complex instrument control; 52 motorized functions needed to automate the instrument are actually 52. Most of them are symmetrical functions, identically mirrored on the four channels. The symmetry is not present only in the management of IFU. The electro-mechanical subsystems to be controlled are:

- Filter Exchange Unit (FEU)
- Mask Exchange Unit (MEU)

- Grism Exchange Unit (GEU)
- Mask Shutters Unit (MSU)
- Focusing Unit (FU)
- 1 Integral Field Unit (IFU)

Each unit is a complex electro-mechanical system, requiring many motorized functions.

The FEU is a 10 filter juke-box composed of a linear filter selector and a filter translator which moves the selected filter inside the optical path.

The MEU is a 15 mask exchanger equipped with a linear mask selector, a mask clamp to bring the selected slit mask, a mask translator to carry the mask to the focal plane, and a mask blocker to block the mask in position.

The GEU is a 6 grism exchanger, composed of a grism selector (a rotating carousel), a grism translator to move the selected grism to the optical path and a grism blocker to block the grism in the right position.

The MSU is composed of a couple of curtains able to obscure a selectable part of the field for calibration purposes.

The FU moves a lens in the camera in order to obtain the best focus.

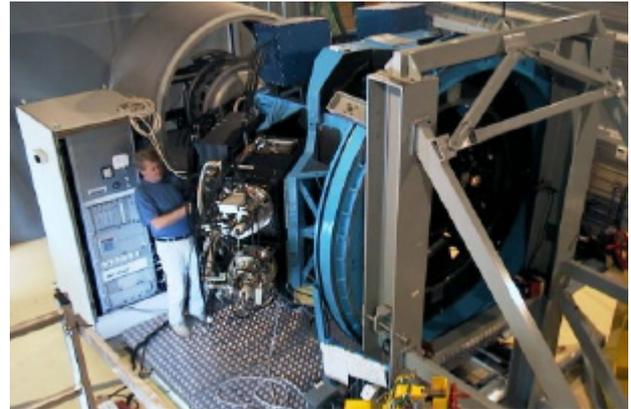

Figure 3: VIMOS in assembling phase at Observatoire de Haute Provence

The IFU is composed, from the control point of view, of an elongator, able to change the magnification; a shutter, able to reduce the incoming light; four masks which are moved in pairs from the parking position to the focal plane when needed.

### 2.2 Instrument functions control

The specification requirements set to 90 sec the maximum time to change the total configuration of the instrument. This means that all the electro-mechanical subsystems (Filter Exchange Unit, Grism Exchange Unit, Mask Exchange Unit, Mask Shutters, Focusing Units, Integral Field Unit) must work in parallel without cross-interferences.

The motions requiring a fine position control are implemented using stepper motors with internal counting position measurement (therefore no encoders are used).

The motions requiring only a movement between two fixed positions are implemented using DC motors in speed control mode with two hardware limits.

In case some motions are mutually exclusive, they are hardware interlocked.

Besides the coordination and control of each of the 52 motions, other minor functions (beam temperature monitoring, cabinet cooling system management, calibration lamps management, housekeeping control) are performed to guarantee the proper operation of the instrument.

### 2.3 Low level software

A low level software library called Device Test Software (DTS) has been created to drive the instrument electronics and tune the operation of the instrument. It has been realized by the OAC team and integrated inside the overall Instrument Control Software (ICS) developed by the Observatoire Midi Pyrénées team. This library drives the electro-mechanical functionalities of the instrument, allowing the higher level software to access and control them without taking care of the electro-mechanical details.

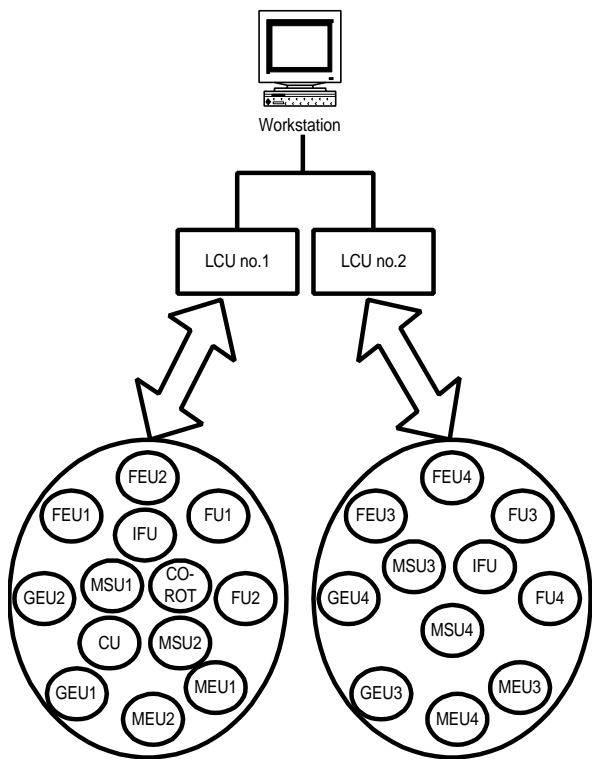

Figure 2: Subsystems division between the Local Control Units

## 2.4 Hardware architecture

The low level software runs on two VME Local Control Units, hosted in two separate control cabinets, each equipped with a Motorola MVME167 CPU card. The division of tasks in two LCUs was an obliged choice due to the high number of systems to be controlled. The operating system is WindRiver VxWorks; the VME crates are equipped with VME ESO standard boards. The two LCUs are managed by an HP-UX workstation. All the VME boards and electronic components have been chosen in conformity with ESO VLT instrumentation standards, in order to ensure compatibility with other ESO applications and simplify maintenance and integration in the VLT environment.

In each LCU the stepper motors are controlled in position mode by four pairs of Maccon Mac4-stp controller + ESO VME4ST amplifier boards, and the DC motors are controlled in speed mode by three ESO VME4SA boards.

It is evident that the two LCUs are highly overloaded with motor control tasks with respect to other similar applications. Some multitasking problems have arisen while testing the instrument in Europe at the Observatoire de Haute Provence, where the problems have been identified and circumvented.

Temperature data are collected from PT100 sensors to a Ester station via a 4-20mA converter, providing information used for automatic focusing to the LCUs via one of the serial ports of an Esd ISER8 board.

The calibration lamps are managed by digital I/O lines through Acromag boards; other diagnostic lines are used to monitor the status of the system.

The cooling system of the control cabinets is ensured by an ESO compliant cooling controller, communicating with the LCU via a serial port of the ISER8 board.

A large cable co-rotator has been necessary in order to drive and protect the high number of cables during the rotation of the instrument at the Nasmyth focus of the VLT. The motion of the co-rotator is slaved to the telescope instrument rotator through a potentiometer which detects the differential motion and consequently drives the co-rotator, so taking it continuously aligned with the Nasmyth rotator. A relay chain is implemented in order to inhibit the motion of both Nasmyth rotator and cable co-rotator in some emergency conditions (emergency keylock, limit switches, etc.).

## 3 ACKNOWLEDGEMENTS

The authors wish to thank O. Le Fèvre, G. Vettolani and all the VIRMOS consortium staff for their collaboration along the whole duration of the project.